\documentclass[%
reprint,
superscriptaddress,
showpacs,preprintnumbers,
 amsmath,amssymb,
aps,
prx,
]{revtex4-1}

\usepackage{lineno}

\usepackage{comment}

\usepackage{txfonts}
\usepackage{amsfonts}
\usepackage{color}%
\usepackage{graphicx}
\usepackage{dcolumn}
\usepackage{bm}

\begin{document}

\title{Electrides as a New Platform of Topological Materials}

\author{Motoaki Hirayama}%
\affiliation{%
 Department of Physics, Tokyo Institute of Technology, 2-12-1 Ookayama, Meguro-ku, Tokyo 152-8551, Japan
}%
\affiliation{%
RIKEN Center for Emergent Matter Science, Wako, Saitama 351-0198, Japan
}%

\author{Satoru Matsuishi}%
\affiliation{%
Materials Research Center for Element Strategy, Tokyo Institute of Technology, 4259 Nagatsuta-cho, Midori-ku, Yokohama 226-8503, Japan
}%

\author{Hideo Hosono}%
\affiliation{%
Materials Research Center for Element Strategy, Tokyo Institute of Technology, 4259 Nagatsuta-cho, Midori-ku, Yokohama 226-8503, Japan
}%

\author{Shuichi Murakami}%
\affiliation{%
 Department of Physics, Tokyo Institute of Technology, 2-12-1 Ookayama, Meguro-ku, Tokyo 152-8551, Japan
}%
\affiliation{%
Materials Research Center for Element Strategy, Tokyo Institute of Technology, 4259 Nagatsuta-cho, Midori-ku, Yokohama 226-8503, Japan
}%

\date{\today}

\begin{abstract}
Recent discoveries of 
topological phases realized in electronic states in solids have revealed an important role 
of topology, which ubiquitously appears in various materials in nature.
Many well-known materials have turned out to be topological materials, and this new viewpoint of topology 
has opened a new horizon in material science. 
In this paper we find that electrides are
suitable for achieving various topological phases, including topological insulating and topological semimetal phases.
In the electrides, in which electrons serve as anions,
the bands occupied by the anionic electrons lie near the Fermi level,
because the anionic electrons are weakly bound by the lattice.
This property of the electrides is favorable for achieving 
band inversions needed for topological phases,
and thus the electrides are prone to topological phases.
From such a point of view, we find many topological electrides,
Y$_2$C (nodal-line semimetal (NLS)), Sc$_2$C (insulator with $\pi$ Zak phase), Sr$_2$Bi (NLS), HfBr (quantum spin Hall system), and LaBr (quantum anomalous Hall insulator), by using \textit{ab initio} calculation.
The close relationship between the electrides
and the topological materials is useful in material science in both fields.

\end{abstract}

\maketitle

\section{INTRODUCTION}
In recent years,  a number of topological materials such as topological insulators~\cite{Kane05a,Kane05b,BZ} and 
topological semimetals~\cite{Murakami07b,Mullen15} have been found by various theoretical and  experimental approaches.
Both in the topological insulator phases and in the topological semimetal 
phases, band inversions in the $\bm{k}$-space are required.
To search for materials with band inversions toward discoveries of new topological materials, 
it is customary to look for materials composed of heavy elements with large spin-orbit coupling such as Bi~\cite{Murakami06,Fu07,Zhang09,Wang12,BYan-review,Ando13}.
In particular, search for topological insulators~\cite{Fu07,Zhang09,Hasan10,Qi11},
quantum anomalous Hall systems~\cite{Liu08,Yu10,Chang13},
and Weyl semimetals~\cite{Wan11,Liu14,Hirayama15,Weng15}
is based on largeness of the spin-orbit coupling.
Nevertheless, it is sometimes difficult to design systems with band inversion, particularly spinless systems, i.e., systems with negligible spin-orbit coupling. It is because there are almost no criteria for systematic search for 
topological materials in spinless systems.

In the present paper, we propose that electrides 
are suitable for achieving various types of topological materials, both in spinful and spinless cases.
In electrides, some electrons reside in the interstitial regions, and they are not constrained by the electric field from the nuclei; 
instead, they are present in the valleys of the electric potential from the cations~\cite{Ellaboudy83,Singh93,Sushko03,Matsuishi03}.
Thus, the interstitial electrons are not strongly stabilized by the electric field from the cations, as compared with atomic orbitals, which are stabilized by the strong electric field near the nuclei.
Therefore, the work function in electrides generally becomes small~\cite{Singh93,Toda07}.
As a result, the interstitial states may appear as occupied states located near the Fermi level, 
and it is suitable for realization of band inversion, as compared with materials described well with atomic orbitals.
This concept is common with the recent theories on symmetry indicators
~\cite{Bradlyn17,Po17}, where the electron distribution away from the atomic limit is
the key to topological phases.
Moreover, we show that in some of such topological elecrides,
the interstitial floating states are the topological surface states,
which are the hallmark for some classes of topological materials. 
Spatial distribution of the surface states of topological electrides is quite different from that in conventional topological materials~\cite{Bernevig06,Zhang09}. 
The surface states in the conventional topological materials are composed of atomic orbitals, while those in topological electrides are floating, and there are no atoms at the center of the floating surface states.
The examples of topological electrides in this paper show that the band inversion can be achieved by 
bringing the states near the Fermi level by using anionic electrons.
This viewpoint 
is useful for material design of topological materials.
It is distinct from other 
viewpoints of search and design of topological materials
proposed previously~\cite{Murakami-gapclosing,Gibson15,Bradlyn17,Po17}.

The electrides have several nontrivial material and chemical properties.
One of the characteristic properties of electrides is a low work function, which is 
vital for applications as electronic materials and in chemical reactions.
The concept of utilizing the low work function of electrides in this paper is
similar to that of heavy elements in topological insulators, because
heavy elements which are often used in topological insulators are
have low work functions, being favorable for band inversions. 
Another remarkable property of electrides is that
the two-dimensional electron gas (2DEG) with high electron mobility 
is realized as bulk states in two-dimensional electrides.
It is in contrast with the usual 2DEG, which is realized only at the interface between different semiconductors.
In the topological electrides proposed in this paper, 
the floating topological surface state might realize 2DEG with high mobility and topological robustness.

In order to demonstrate the usefulness of electrides for the design of the topological materials,
we search for topological electrides using first-principles calculation.
The calculations are based on the generalized gradient approximation (GGA) of the density functional theory (DFT).
Since Sc$_2$C to be discussed later contains localized $3d$ orbitals of Sc,
the calculation of Sc$_2$C is done by the GW approximation (GWA) beyond the DFT/GGA~\cite{Aryasetiawan98}.

Indeed, we find that Y$_2$C is a topological nodal-line semimetal (NLS).
Thanks to the bulk nontrivial topology and the properties as an electride, its topological surface states float on the material surface.
We also study Sc$_2$C having the same structure as Y$_2$C,
and find that Sc$_2$C is in a topological insulating phase characterized by the $\pi$ Zak phase,
which leads to quantized nonzero polarization.
Finally, we also introduce other examples of topological electrides, Sr$_2$Bi, HfBr, and LaBr, 
showing a nodal-line semimetal, a quantum spin Hall phase, and a quantum anomalous Hall phase, respectively.
Through these examples, we conclude that electrides are suitable for realizing band inversions, which 
are necessary for various nontrivial topological phases.
Furthermore, the topological surface states coming from these topological phases are floating states in the electrides, 
and they do not reside on atoms; therefore with the surface probes such as scanning tunneling microscope (STM) they are seen quite differently
differently from conventional surface states.

\section{RESULTS}
\subsection{Electronic band structure of Y$_2$C}

\begin{figure*}[htp]
\includegraphics[width=15cm]{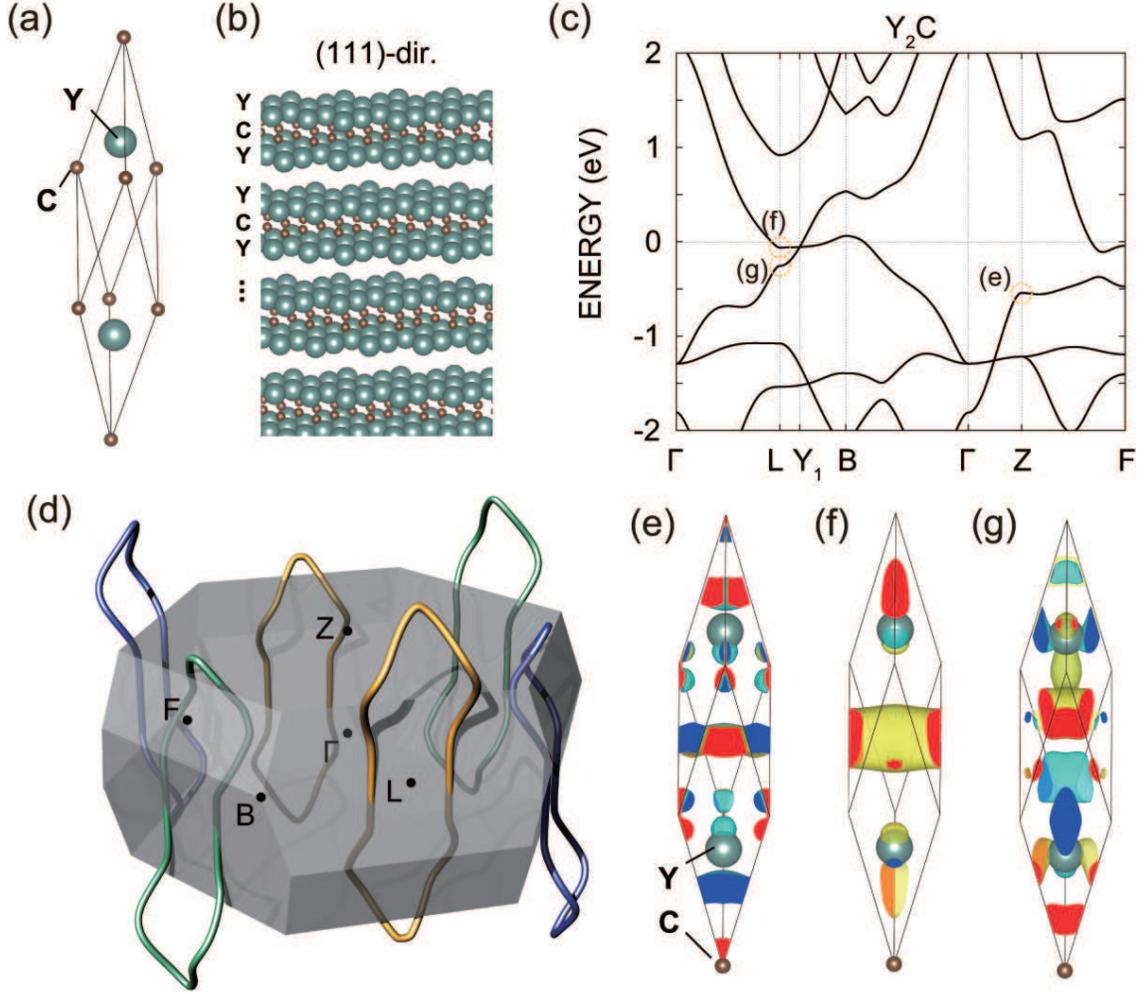}
\caption{\label{fig:Y2C} 
Topological nodal-line in Y$_2$C.
(a) Crystal structure of Y$_2$C and Sc$_2$C.
The blue and brown balls represent Y/Sc and C atoms, respectively.
(b) Layered crystal structure along (111) in Y$_2$C and Sc$_2$C.
(c)  Electronic band structure of Y$_2$C in the generalized gradient approximation (GGA).
The energy is measured from the Fermi level.
(d) Nodal lines and the Brillouin zone of Y$_2$C, where identical nodal lines (modulo the reciprocal vectors) are shown in the same color. In (e), (f) and (g) we
show wavefunctions of some of the eigenstates, with the colors indicate the amplitudes 
of the wavefunctions themselves.  
The yellow and light blue represent the positive and negative components of the wave function, respectively,
and the red and navy blue represent their cross sections, respectively.
(e) Eigenstate of the top of the valence band at the $Z$ point.
(f,g) Eigenstates at $-0.062$ and at $-0.256$ eV at the $L$ point, respectively.
These eigenstates are marked in the band structure in (c).
}
\end{figure*}

Y$_2$C is a layered electride~\cite{Inoshita14,Zhang14}.
Figure~\ref{fig:Y2C}(a) shows the unit cell of Y$_2$C, containing two Y atoms and one C atom.
The crystal forms a layered structure along the (111) plane as shown in Fig.~\ref{fig:Y2C}(b).
It is the same structure as Ca$_2$N known as a two-dimensional electride~\cite{Lee13}.
The space group of Y$_2$C is $R\bar{3}m$ (No. 166), having 
a spatial inversion symmetry, a rotational symmetry $C_3$ about the $z$-axis and a mirror symmetry with respect to the plane perpendicular to the $x$-axis.
Experimentally, Y$_2$C shows metallic behavior~\cite{Zhang14},
and it is nonmagnetic down to at least 2 K~\cite{Zhang14}.

Figure~\ref{fig:Y2C}(c) shows the electronic band structure near the Fermi level of Y$_2$C, 
with the corresponding Brillouin zone shown in Fig.~\ref{fig:Y2C}(d).
The valence band originating from the C $2p$ orbitals and the conduction band originating from the Y $4p$ orbitals exist near the Fermi level.
In addition to these, a band originating from interstitial electrons exists at the top of the valence band (Fig.~\ref{fig:Y2C}(e))~\cite{Dale17};
which is also the case in the electride Ca$_2$N~\cite{Zhao14}.
This band is almost occupied, and provides two electrons in 
the space between the Y$_2$C layers surrounded by cations of Y, 
stabilizing the crystal structure in the form of [Y$_2$C]$^{2+}2e^{-}$.
At the $L$ point, being one of the time-reversal invariant momenta (TRIM), bands are inverted between the eigenstates corresponding to the interstitial electrons and those originating from the Y $4d$ orbitals (Figs.~\ref{fig:Y2C}(f,g)).
As a result, a degeneracy between the valence and the conduction bands appears at $Y_1$ on the $L$-$B$ line.
This degeneracy on the L-B line is protected by the $C_2$ symmetry around the $x$ axis.
With respect to the $C_2$ rotation whose axis goes through the interstitial region, the $s$-like interstitial orbital is even (Fig.~\ref{fig:Y2C}(f)). Meanwhile, the eigenstate originating from the Y $4d$ orbital is odd because its amplitude has 
the opposite signs between the neighboring Y$_2$C layers (Fig.~\ref{fig:Y2C}(g)).
For this reason, the interstitial band and the band originating from the Y $4d$ orbital do not hybridize with each other on the $L$-$B$ line and have the degeneracy 
at the band crossing.
We note that within the GWA, Y$_2$C has a narrow gap as shown in Appendix A.
Nevertheless, the result with the GGA in Fig.~\ref{fig:Y2C}(a) is considered to be more precise,
and it agrees with \cite{Huang18}.

In fact, this degeneracy appears not at the isolated point $Y_1$ on the $L$-$B$ line, but extends along a loop in ${\bf k}$ space, when the spin-orbit interaction (SOI) is neglected, 
as has been found in \cite{Huang18}. Therefore Y$_2$C is a NLS.
This follows because of the time-reversal and the inversion symmetry~\cite{Hirayama17}.
In Y$_2$C, nodal lines (NLs), along which the valence band and the conduction band are degenerate, encircle
around the $L$ points, as shown in Fig.~\ref{fig:Y2C}(d).
There are three NLs, which are transformed to each other by the $C_3$ symmetry.
Although Y$_2$C is known as a two-dimensional electride, in which some electrons are distributed between the layers,
the interstitial band is dispersive in the $k_z$-direction, and hereby the NLs are of three-dimensional nature.
These nodal lines originate from the $\pi$ Berry phase, 
as we
directly confirmed by the calculation of the Berry phase around the NL,
\begin{align}
\phi (\ell)=-i\sum_{n}^{\text{occ.}}\int_{\ell} d {\bf k}\cdot 
\left\langle{u_n({\bf k})}\right| \nabla_{\bf k} \left|{u_n({\bf k})}\right\rangle,
\label{eq:phi}
\end{align}
where $u_n({\bf k})$ is the bulk eigenstate in the $n$-th band,
the sum is over the occupied states, and the system is assumed to have a gap
everywhere along the loop $\ell$ \cite{Hirayama17,Hirayama18JPSJ}. 
It is similar to those in the alkaline earth metals Ca and Sr \cite{Hirayama17}.
This type of NLs appear in spinless systems with the time-reversal and the inversion symmetries, and the material Y$_2$C belongs to
this class.

\subsection{Topological surface states and Zak phase}

\begin{figure}[htp]
\includegraphics[width=8cm]{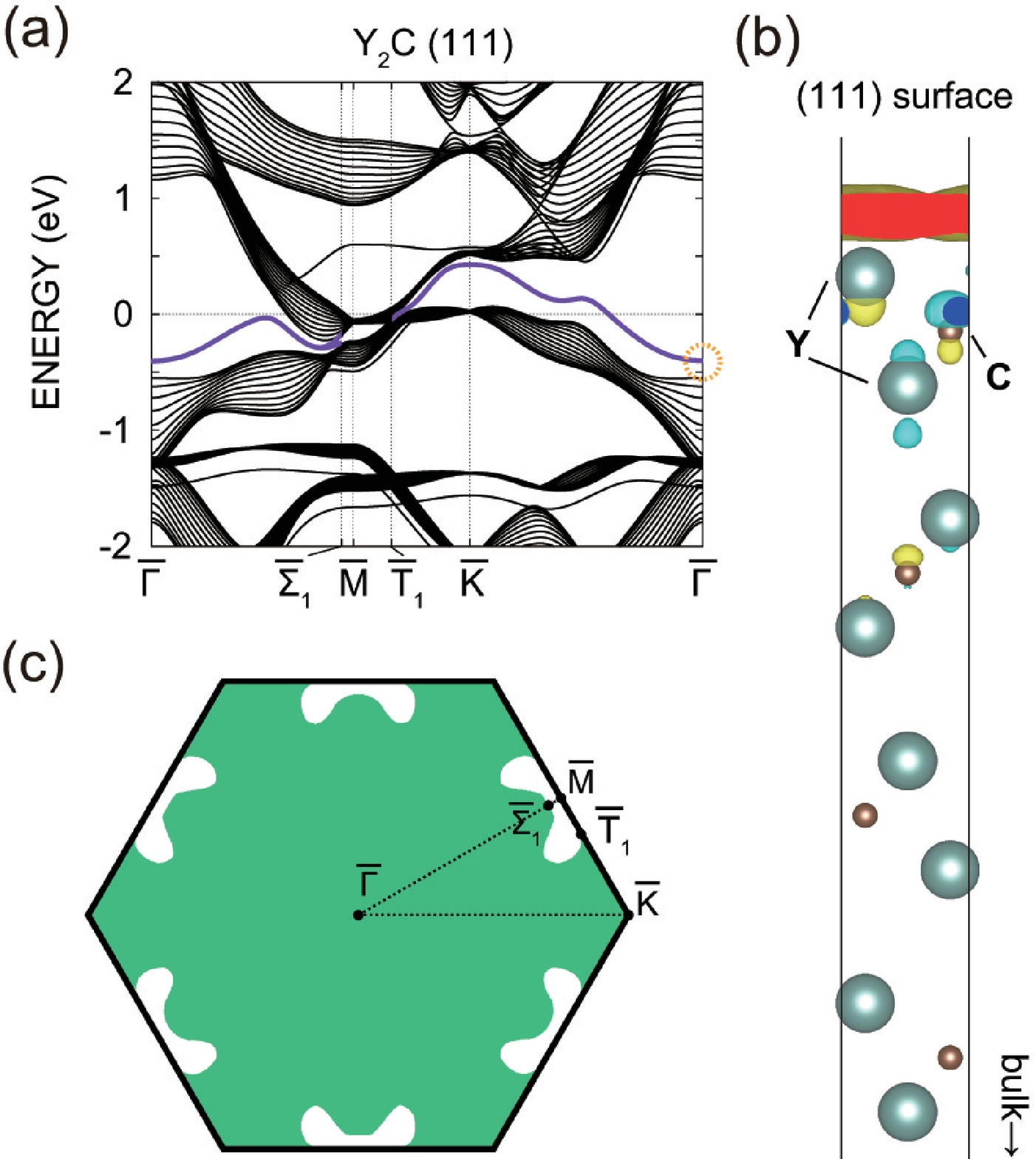}
\caption{\label{fig:Y2Csf} 
Floating topological surface states in Y$_2$C.
(a) Electronic band structure of the slab of Y$_2$C with surfaces along (111) in the GGA.
The symmetry points are $\bar{\rm \Gamma} =$(0, 0, 0), $\bar{\rm X}=(\pi/2\bar{\rm a})$(0.5, 0.5, 0), and $\bar{\rm K}=(\pi/2\bar{\rm a})$(1, 0, 0), where $\bar{\rm a}$ is the lattice constant for the surface.
The energy is measured from the Fermi level.
The purple band represents the topological surface states extending from the nodal line near the Fermi level.
(b) Charge distribution of the drumhead surface state at the $\Gamma $ point.
The corresponding point is indicated by the dotted circle in (a).
The yellow and light blue represent the positive and negative components of the wave function, respectively,
and the red and navy blue represent their cross sections, respectively.
(c) Dependence of the Zak phase on the surface momentum ${\bf k}_{\parallel}$. The shaded region represents ${\bf k}_{\parallel}$  with the $\pi$ Zak phase, while other regions represent that with the $0$ Zak phase. 
}
\end{figure}

Next, we show the result of the calculation of the surface states.
Figure~\ref{fig:Y2Csf}(a) shows the electronic band structure of a 15-layer slab of Y$_2$C.
Drumhead surface states~\cite{Kim15,Yu15,Hirayama17}, which are characteristic surface states in nodal-line semimetals, 
emerge in the region surrounded by the projection of NLs.
In this figure, the $\bar{\Sigma}_1$ and $T_1$ points are the positions of the NLs projected onto the two-dimensional Brillouin zone,
from which the drumhead surface states extend.
In the angle-resolved photoemisison spectroscopy (ARPES) measurements~\cite{Horiba17}, 
there might be some signal coming from the surface band in the intersection of the valence and conduction bands,
which might be the above drumhead surface states.
Figure~\ref{fig:Y2Csf}(b) shows the spatial distribution of a wavefunction in the drumhead surface states at the $\bar{\Gamma}$ point.
The surface states have characteristic spatial distributions as an electride; they do not localize around the surface atoms but mainly appears as floating states on the surface.
Such spatial distribution of the floating surface states can be detected with the STM with a spatial resolution in the depth direction, in a way similar to the STM measurement on TaAs~\cite{Inoue16}.

Here we note that the NLs are closely related to the Zak phase.
The Zak phase is defined by the integral of the Berry connection along a reciprocal
lattice vector ${\bf G}$ in the reciprocal space. 
To define the Zak phase, we decompose the wavevector ${\bf k}$ into the component
along ${\bf n}\equiv {\bf G}/|{\bf G}|$ and those perpendicular to ${\bf n}$:
${\bf k}=k_{\perp}{\bf n}+{\bf k}_{\|}$, ${\bf k_{\|}}\perp{\bf n}$. 
Then, for each value of  ${\bf k}_{\parallel} $, 
the Zak phase is defined by
\begin{align}
\theta ({\bf k}_\parallel )=-i\sum_{n}^{{\rm occ.}}\int _{0 }^{|{\bf G}|} d k_{\perp} 
\left\langle{u_n({\bf k})}\right| \nabla_{k_{\perp}} \left|{u_n({\bf k})}\right\rangle ,
\label{eq:Zak}
\end{align}
where 
$u_n({\bf k})$ is the periodic part of the bulk Bloch wavefunction in the $n$-th band, with the gauge choice
$u_n({\bf k})=u_n({\bf k}+{\bf G})e^{i{\bf G}\cdot{\bf r}}$, 
and the sum is over the occupied states.
Figure~\ref{fig:Y2Csf}(c) shows the distribution of the Zak phase for each $\bm{k}_{\|}$ point in the two-dimensional Brillouin zone along the (111) plane. The Zak phase is $\pi$ in the shaded region and is zero in the blank region, and these two regions are divided by the 
projection of the nodal lines onto the (111) plane. Because
we neglect the spin-orbit coupling, the Zak phase is quantized to be zero or $\pi$
(mod $2\pi$) due to the time-reversal and inversion symmetries.

These $\pi$ / $0$ values of the Zak phase corresponds to
the presence/absence of midgap boundary states, which are
similar to those in the Su-Schrieffer-Heeger (SSH) model \cite{SSH}. This can be easily confirmed by
a comparison between Figs.~\ref{fig:Y2Csf}(a) and (c).
As shown for calcium in \cite{Hirayama17}, at the wavevector $\bm{k}_{\|}$ having the $\pi$ Zak phase, there appear surface polarization charges
equal to $\frac{e}{2}$ (mod $e$), if the system is regarded as a one-dimensional insulator at the given ${\bf k}_{\|}$. 
The area with the $\pi$ Zak phase is 90.4 $\% $ of the total area of the Brillouin zone.
Therefore, Y$_2$C is close to the topological transition into an insulator with $\pi$ Zak phase for every ${\bf k}_{\|}$

The surface states  in 
our result (Fig.~\ref{fig:Y2Csf}(a)) cover almost the entire surface Brillouin zone.
As we mentioned, the distribution of the Zak phase 
in Fig.~\ref{fig:Y2Csf}(c) is consistent with this surface state distribution. Furthermore,
the values of the Zak phase at the surface
TRIM can be calculated from the parity eigenvalues of the bulk eigenstates as we explain 
in Appendix B, and are consistent with Fig.~\ref{fig:Y2Csf}(c). 
In particular, the Zak phase at $\bar{\Gamma}$ is determined to be $\pi$ (see Fig.~\ref{fig:Y2Csf}(c)) 
from the fact that the products of the parity eigenvalues over the occupied bands 
are inverted between the $\Gamma$ and the $Z$ points.
This parity inversion between $\Gamma$ and $Z$ physically originates from the 
occupied interstitial states. Therefore, this
appearance of the topological surface states at $\bar{\Gamma}$ is a manifestation
of the physics of electrides, leading us to the idea of ``topological electrides''. 
These features at $\bar{\Gamma}$ are missing in 
the previous work \cite{Huang18} (see Appendix B).

\subsection{Sc$_2$C as an Insulator with the $\pi$ Zak phase}

\begin{figure*}[htp]
\includegraphics[width=15cm]{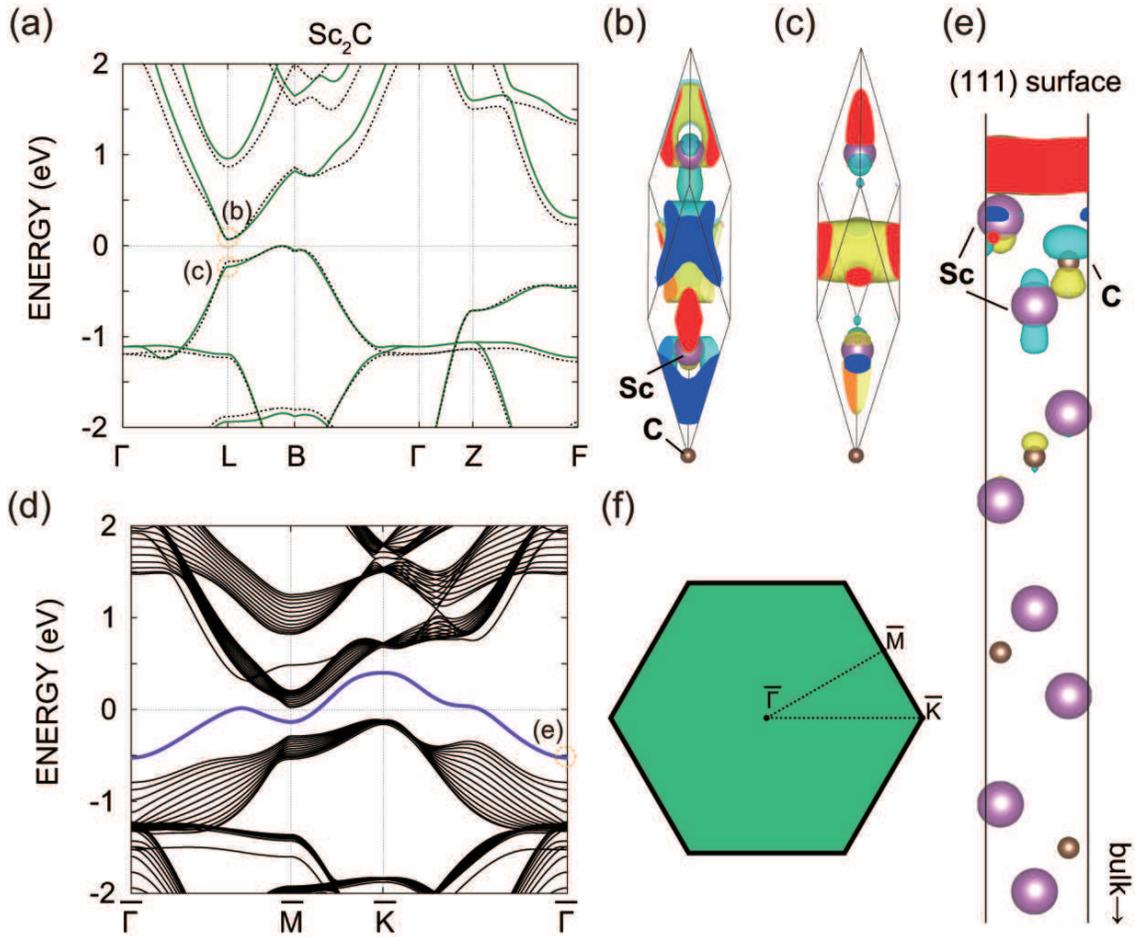}
\caption{\label{fig:Sc2C} 
Sc$_2$C as an insulator with $\pi$ Zak phase.
(a) Electronic band structure of Sc$_2$C.
The solid green line is the result in the GWA and the black dotted line is the result in the GGA+U, where $U$ is adjusted to reproduce the band gap of the GWA.
(b, c) Wavefunctions of the eigenstates at the bottom of the conduction band and that at the top of the valence band at the $L$ point, respectively.
The corresponding points are indicated by the dotted circles in (a).
The yellow and light blue represent the positive and negative components of the wave function, respectively,
and the red and navy blue represent their cross sections, respectively.
(d) Electronic band structure for the (111) surface of a slab of Sc$_2$C in the GGA+U.
The purple band represents the topological surface state originating from the quantized $\pi$ Zak phase.
(e) Wavefunction of the topological surface state at the $\bar{\Gamma} $ point.
The corresponding point is indicated by the dotted circle in (d).
(f) Dependence of the Zak phase on the surface momentum ${\bf k}_{\parallel}$.
The shaded region represents ${\bf k}_\parallel$ with the $\pi$ Zak phase.
In contrast with Fig.~\ref{fig:Y2Csf}(c),
the Zak phase is $\pi$ for the entire two-dimensional Brillouin zone. 
The energy is measured from the Fermi level.
}
\end{figure*}

Next, we show the result of Sc$_2$C.
Sc$_2$C is predicted to be a stable electride with the same structure as Y$_2$C in the \textit{ab initio} calculation~\cite{Zhang17}.
Figure~\ref{fig:Sc2C}(a) shows the band structure of Sc$_2$C in the GWA~\cite{Aryasetiawan98}.
Since the Sc $3d$ orbitals are more localized compared to the Y $4d$ orbitals,
Sc$_2$C is an insulator due to the decrease in bandwidth and the increase in correlation effects.
Actually, the energies of the eigenstates at the $L$ point around the Fermi level are reversed from those in  Y$_2$C (Figs.~\ref{fig:Sc2C}(b,c)).

Next, we calculate the Sc$_2$C surface states with the GGA+U where the value of $U$ is chosen to reproduce the band gap in the GWA (Fig.~\ref{fig:Sc2C}(a)).
Figure~\ref{fig:Sc2C}(d) shows the electronic band structure of a 15-layer slab of Sc$_2$C.
There appear surface states in the gap, and they are separated from the bulk states.
As is similar to Y$_2$C, the surface states of Sc$_2$C at the $\Gamma $ point are not localized at the surface atoms but are floating on the surface (Fig.~\ref{fig:Sc2C}(e)).
The wavefunctions near the $M$ point originate from the Sc $3d$ orbitals due to hybridization between the floating states and the Sc $3d$ orbitals.
We find that the Zak phase along the (111) direction is equal to $\pi$ in the entire Brillouin zone (Fig.~\ref{fig:Sc2C}(f)).
When the Zak phase is $\pi$ at every $\bm{k}_{\|}$, 
there are surface polarization charges equal to $e/2$ (mod $e$) per a surface unit cell 
on the (111) surface  \cite{Hirayama17}, and because of the inversion symmetry, 
the total surface charge (due to bulk bands) is $e$ (mod $2e$).
Thus, the number of occupied bands in the bulk is changed by an odd number from that in the case of the $0$ Zak phase.
Therefore, it is necessary to compensate the deficiency of the bulk charges somewhere when the Zak phase is $\pi$.
If the deficiency is compensated by the bulk states, the energy cost is proportional to the number of bulk layers; hence, it is highly improbable when there
is a sizable band gap.
On the other hand, when the deficiency is compensated by the surface, the energy cost is not dependent on the number of the bulk layers.
Therefore, in a sufficiently thick slab, due to the topological requirement from the Zak phase together with the above energetic reason, topological surface states 
inevitably appears in the gap to accommodate the deficiency of the bulk charges as
demonstrated in Fig.~\ref{fig:Sc2C}(d).
For consistency, we checked the stability of the obtained electronic structure using a doubled unit cell,
and confirmed that the change of the surface structure with a longer period is not seen.
Thus we have confirmed that the stable topological surface state emerges without dimerization.

\begin{figure}[htp]
\includegraphics[width=8cm]{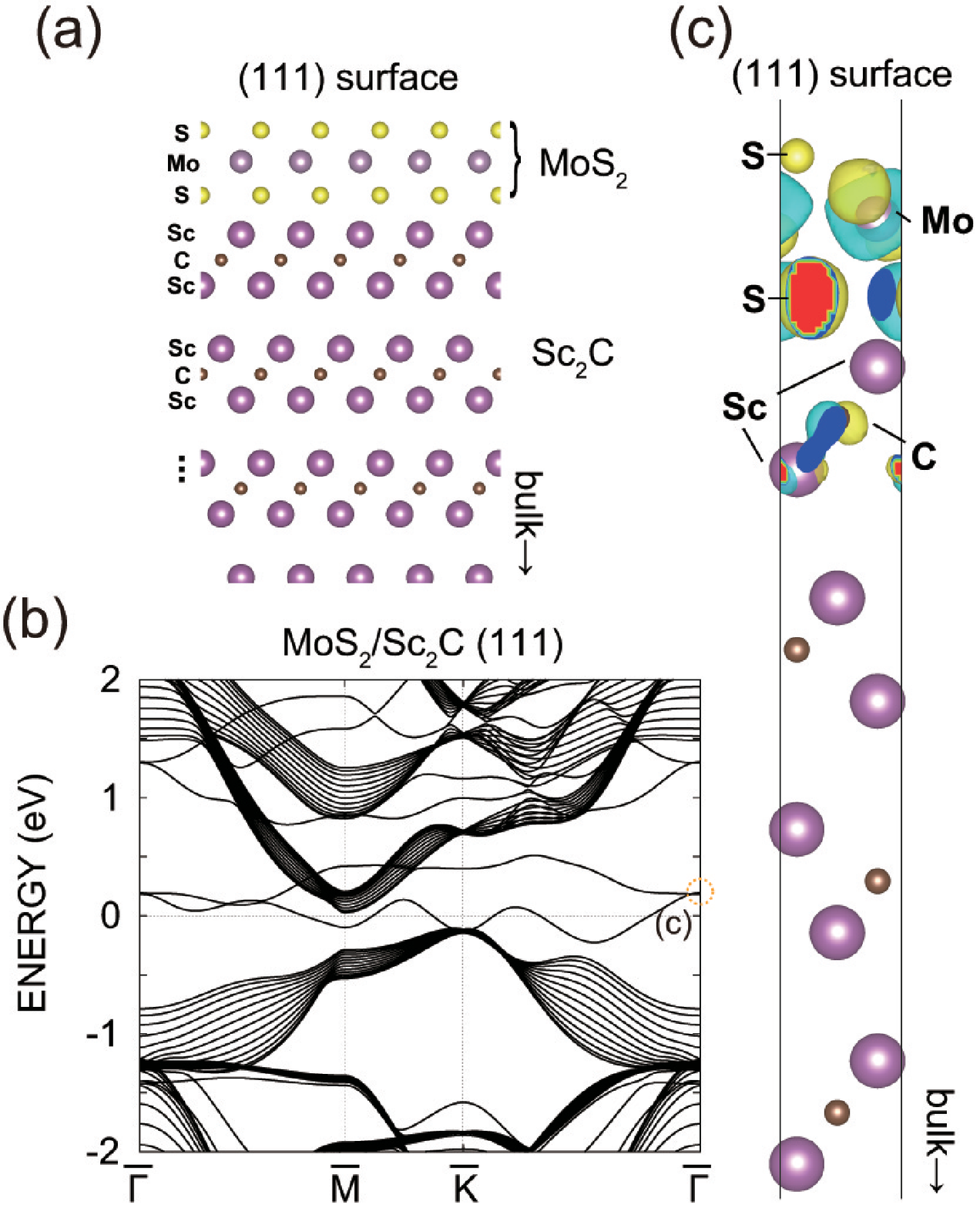}
\caption{\label{fig:dope} 
Doping via topological surface charges of Sc$_2$C.
(a) Crystal structure of MoS$_2$ on the Sc$_2$C (111) surface.
(b) Electronic band structure of a MoS$_2$ monolayer on Sc$_2$C.
(c) Wavefunction of the eigenstate at the $\Gamma $ point originating from the MoS$_2$ monolayer.
The corresponding point is indicated by the dotted circle in (b).
The energy is measured from the Fermi level.
}
\end{figure}

Next, we propose a method of huge carrier doping using the topological 
surface charges in Sc$_2$C.
As an example, we consider a system in which a MoS$_2$ monolayer is placed on a 15-layer slab of Sc$_2$C (Fig.~\ref{fig:dope}(a)).
The lattice constant of MoS$_2$ is close to that of the Sc$_2$C substrate ($a=3.323$\AA\ in Sc$_2$C and $a=3.161$\AA\ in MoS$_2$).
MoS$_2$ is a well-known layered insulator and has no polarization in the bulk~\cite{Mak10,Splendiani10}.
By the same reason given in the previous paragraph, 
as long as the symmetry does not change, the interface between an insulator with the 0 Zak phase and one with the $\pi$ Zak phase is necessarily metallic.
Since the topological surface states of Sc$_2$C are not localized around the atoms but they float, they can be easily transferred to MoS$_2$.
Figures~\ref{fig:dope}(b,c) are the electronic band structures of  the MoS$_2$ monolayer on Sc$_2$C and the eigenstates.
We can see that huge carrier doping equivalent to one electron per unit cell is achieved without elemental substitution, 
because a half of the bands at the Fermi level originating from the Mo $d_{x^2-y^2}$ and $d_{xy}$ orbitals of MoS$_2$ are occupied.
As a comparison, in the carrier doping by the field effect transistor (FET), which does not change the composition, the number of doped carriers is at most $1\%$ per unit cell.
Thus this topological carrier-doping is versatile since it does not require direct coupling between the target material and the Sc$_2$C layer.

We also consider electron doping into a molecule by using Sc$_2$C
and show the result in Appendix C.
Especially, we study a system  with PH$_3$ (phosphine) molecules on the Sc$_2$C (111) surface,
where the carriers are doped to hydrogen atoms at atmospheric pressure.
Carrier doping to hydrogen atoms would open a way to high-temperature superconductivity originating from phonons at hydrogen atoms~\cite{Li14,Drozdov15}.

\subsection{Various topological electrides}

There are various classes for topological materials, depending on 
presence or absence of the SOI and that of the band gap.
So far, physical origins of topological band structures are
discussed mostly based on atomic orbitals. 
In the present paper, we propose that interstitial electrons in electrides are
suitable for realizing topological materials. 
Below, we show the topological degeneracy and band inversion originating from interstitial electrons in realistic materials.
Experimental synthesis is summarized in Appendix D.

\begin{figure}[htp]
\includegraphics[width=8cm]{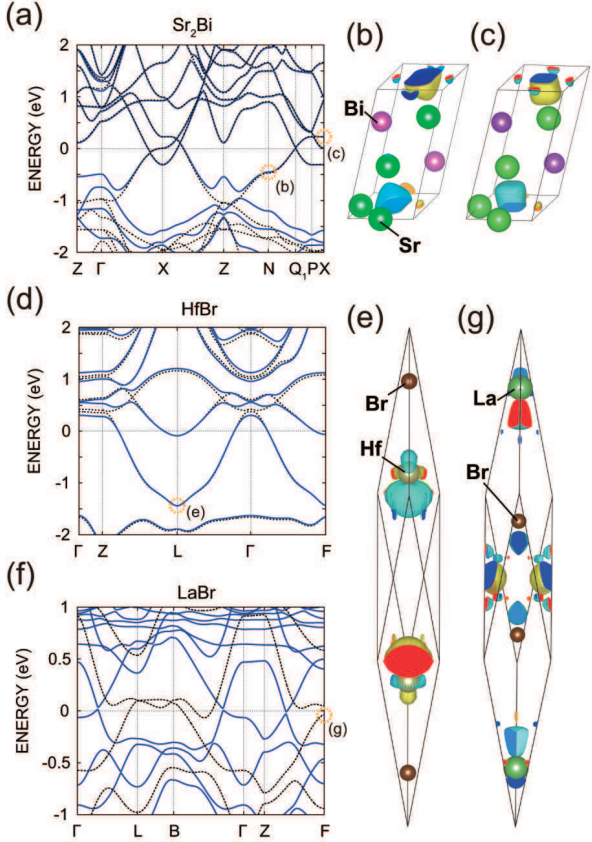}
\caption{\label{fig:mat} 
Band structure of predicted topological electrides.
(a) Electronic band structure of Sr$_2$Bi in the GGA
, where Z$=(0,0,1)$, $\Gamma =(0,0,0)$, $X=(1/2,1/2,0)$, $N=(1/2, 0, 1/2)$, and $P=(1/2, 1/2, 1/2)$ are high-symmetry points in the $\bm{k}$ space.
The blue solid line and the black dotted line show the result without and with the SOI, respectively.
(b, c) Wavefunctions of the eigenstates originating from the interstitial electron at the $N$ and $X$ points in the calculation without the SOI.
(d) Electronic band structure of HfBr in the GGA.
The blue solid line and the black dotted line show the result without and with the SOI, respectively.
(e) Wavefunction of the eigenstate originating from the interstitial electron at the $X$ point in the calculation without the SOI.
(f) Electronic band structure of LaBr in the GGA.
The blue solid line and the black dotted line shows the result without and with the SOI and magnetism, respectively.
(g) Wavefunction of the eigenstate originating from the interstitial electron at the $F$ point without the SOI.
The results shown in (b) and (c), (e), and (g) corresponds to the states indicated by the dotted circles in (a), (d), and (f), respectively.
The Brillouin zones of HfBr and LaBr are same as that of the Y$_2$C.
}
\end{figure}

We first discuss Sr$_2$Bi, whose space group is $I4/mmm$ (No. 139).
Figure~\ref{fig:mat}(a) is the band structure of bulk Sr$_2$Bi.
A thin film of Sr$_2$Bi is predicted to be a two-dimensional NLS~\cite{Niu17}.
Here we find that the bulk Sr$_2$Bi is 
also a NLS, with its NL having  the $\pi$ Berry phase. 
At the $Q_1$ point on the $N$-$P$ line which is a symmetry axis for $C_2$ rotation, two
bands cross, and because of the $\pi$ Berry phase, the NL extends from this $Q_1$ point 
to a general position in $k$ space.
Moreover, we find that
one of the bands constituting the NL originates
not from the atomic orbitals but from
the interstitial electrons, lying between the Sr atoms. 
The top of the valence band is the interstitial states.
Figures~\ref{fig:mat}(b,c) show the interstitial states at the $N$ and $X$ points connecting to the degenerate point $Q_1$.
We also show the effect of the SOI in Fig.~\ref{fig:mat}(a).
The splitting originating from the SOI is large in some $\bm{k}$ region near the Fermi level.
For example, energy splitting over 0.5 eV can be seen at the $Z$ point near $-1$ eV.
Actually, the effect of the SOI in electrides is rarely discussed in the literature, because
the wavefunctions of the interstitial electrons near the Fermi level usually has no node 
and there is no strong electric field from the nuclei.
However, the present case shows that it is possible to induce a large SO splitting in the interstitial bands by hybridization with orbitals having the strong SOI.

HfBr is a layered electride and its space group is $R\bar{3}m$ (No. 166).
The top of the valence band originates from the interstitial electrons existing between the Hf atoms (Figs.~\ref{fig:mat}(d,e)).
The interstitial band and the conduction band originating from the Hf $5d$ orbital at the $\Gamma$ point are inverted by the SOI, making the system a weak topological insulator, i.e. a two-dimensional quantum spin Hall system~\cite{Zhou15}.

LaBr is also a layered electride, with its space group $R\bar{3}m$ (No. 166).
The band originating from the $4f$ orbital of La is in 1-2 eV above the Fermi level, and thus it is not involved in magnetism.
When the spin degree of freedom is ignored, LaBr has a flat band on the Fermi level (blue solid lines in Fig.~\ref{fig:mat}(e)).
Both in the cases with and without the SOI, ferromagnetism develops to eliminate this peak of the density of states at the Fermi level.
A gap then opens at the Fermi level with the SOI, and the system becomes a quantum anomalous Hall insulator~\cite{Wu17,Dolui15}.
When the SOI and the ferromagnetism are considered, the band structure is shown as 
the black dotted lines in  Fig.~\ref{fig:mat}(f). 
Here, one of the bands around the Fermi level originates from the interstitial electrons existing between the La atoms (Fig.~\ref{fig:mat}(g)).

Thus, we have seen that there is a close relationship between electrides and 
topological materials. We have shown this relationship by relating the low work functions
of electrides with band inversions in topological materials. 
We have also pointed out the unique feature of topological electrides: the 
topological floating surface states. These features have not been found in previous 
works discussing on topological band structure in electrides \cite{Zhang18, Huang18, Park18}. 
In particular, in \cite{Zhang18}, the Dirac-type band degeneracy is discussed in the pseudo zero-dimensional 
electride Ca$_3$Pb. Nevertheless, this material is neither a toplogical insulator nor a topological semimetal. Furthermore, it is not obvious how to relate the zero-dimensional nature of the electride to 
the topological nature of the band structure. Next, 
in \cite{Park18}, the electrides Cs$_3$O and Ba$_3$N are found to have
nontrivial band topology. Nevertheless, they are neither topological insulators nor topological semimetals. While Cs$_3$O has a nodal line near the Fermi level, it is not 
a topological semimetal because of the presence of other bands at the Fermi energy. In
Ba$_3$N, the topological band degeneracy is not near the Fermi level.

\section{CONCLUSION}
To summarize, we show that some electride materials such as Y$_2$C and Sc$_2$C 
are topological materials, and some topological materials such as Sr$_2$Bi, HfBr and LaBr
are electrides.
Through these examples, we show that electrides are favorable for
achieving band inversion.
In electrides, because 
the work function is shallow due to the interstitial states, 
various topological phases appear, both in the cases with or without the relativistic effect.
In some classes of topological materials, there appear
surface states coming from the nontrivial topology, and such surface states are
nothing but the floating states from interstitial electrons.

In this study, we used not only the DFT/GGA but also the GWA to accurately calculate the topological characteristics of the electrides.
The topological aspect of the electride is robust as it is protected by the nontrivial nature of the bulk.
Such a topological nature in bulk and surface states could be experimentally detected in various ways.
For example, the floating topological surface states of the insulator with $\pi$ polarization can be detected by the ARPES and the STM.
Topological carrier doping can be detected in experiments such as transport measurements.
Topological carriers in Sc$_2$C can be doped with the concentration of 
$1.05\times 10^{15}$  ($e$ cm$^{-2}$). It is ten times that of the electric double layer (EDL), by which 
$10^{14}$ ($e$ cm$^{-2}$) carrier can be doped to MoS$_2$ monolayer~\cite{Ye12}.
This suggests a possibility of Sc$_2$C as a \textit{topological substrate} for two-dimensional systems and heterostructures.
Thus the combination of the electrides with the concept of topological phases leads us to new possibilities for 
materials science.

\section{METHODS}
The electronic structure and the lattice optimization is obtained from the generalized gradient approximation (GGA) and GGA+U of the density functional theory (DFT).
We use the Perdew-Burke-Ernzerhof (PBE) functional in the GGA~\cite{Perdew96}.
We use  the \textit{ab initio} code OpenMX (http://www.openmx-square.org/) based on localized basis functions and norm-conserving pseudopotentials.
We employ the $6\times 6\times 1$ $\bm{k}$-point sampling for the lattice optimization.
We employ the  $12\times 12\times 12$ and $12\times 12\times 1$ $\bm{k}$-point samplings for the electronic structure of the bulk and the slab, respectively.
Valence orbital set is $s^3p^2d^2$ for Y, $s^2p^2d^1$ for C, $s^4p^3d^2$ for Sc,
$s^3p^1d^2f^1$ for Sr, $s^3p^3d^3f^2$ for Bi, $s^3p^2d^2f^1$ for Hf, $s^3p^3d^2$ for Br, and $s^3p^3d^2f^1$ for La.
The energy cutoff for the numerical integrations is 150 Ry.
The fully relativistic effect including the spin-orbit coupling is calculated in the non-collinear DFT calculations.
The self-energy correction in the GW approximation (GWA)~\cite{hedin65} is obtained from the full-potential linear muffin-tin orbital code~\cite{schilfgaarde06,miyake08}.
We employ the $8\times 8\times 8$ $\bm{k}$-point sampling in the GWA.
The muffintin radii in the GWA (bohr) is 3.1 for Y, 1.55 for C, and 2.75 for Sc. 
The angular momentum cutoff in the GWA is taken at $l = 4$ for all the sites.
The $67\times 2$ unoccupied conduction bands are included in the GWA, where $\times 2$ is the spin degrees of freedom.
For the calculation of the surface in the GGA+U, we determine the value of $U$ to reproduce the result of the GWA.
We introduce empty sphere to match the center of the Wannier function of the interstitial state~\cite{Inoshita17}.
We use 3 eV for the Sc $3d$ orbitals and 1 eV for empty sphere between the Sc$_2$C layers as $U$.
We take 5 layers of Sc$_2$C (15 atoms) for the (001) surface
for the lattice optimization of MoS$_2$ and PH$_3$ structure.
We take 15 layers of Sc$_2$C (45 atoms) for the (001) surface in the calculation of the electronic band structure.

\begin{acknowledgments}
This work was supported by JSPS KAKENHI Grant Number 
 26287062, and by the MEXT Elements Strategy Initiative to Form Core Research
Center (TIES).
HH acknowledges a Grant-in-Aid for Scientific Research (S) (Grant No. 17H0653)
from JSPS.
\end{acknowledgments}


\appendix{}
\section{ELECTRONIC BAND STRUCTURE OF Y$_2$C IN THE GWA}

\begin{figure}[htp]
\includegraphics[width=8cm]{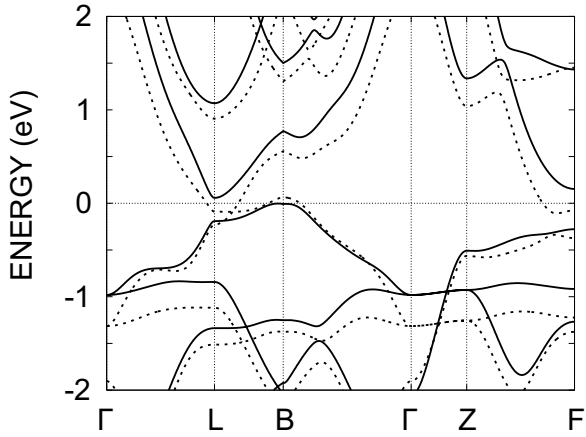}
\caption{
Electronic band structure of Y$_2$C in the GWA.
Electronic band structures of Y$_2$C in the GWA and the LDA. 
The solid line and the dotted line show the results in GWA and those in LDA, respectively.
The energy is measured from the Fermi level.
}
\label{fig:Y2CGW}
\end{figure}

Figure~\ref{fig:Y2CGW} shows the electronic band structure of Y$_2$C in the GWA. 
We employ the $12\times 12\times 1$ $\bm{k}$-point sampling in the LDA
and the $8\times 8\times 8$ $\bm{k}$-point sampling in the GWA.
The $67\times 2$ unoccupied conduction bands are included in the GWA, where $\times 2$ is the spin degrees of freedom.

\section{ZAK PHASE AND PARITY EIGENVALUES IN Y$_2$C}

\begin{figure}[htp]
\includegraphics[width=8cm]{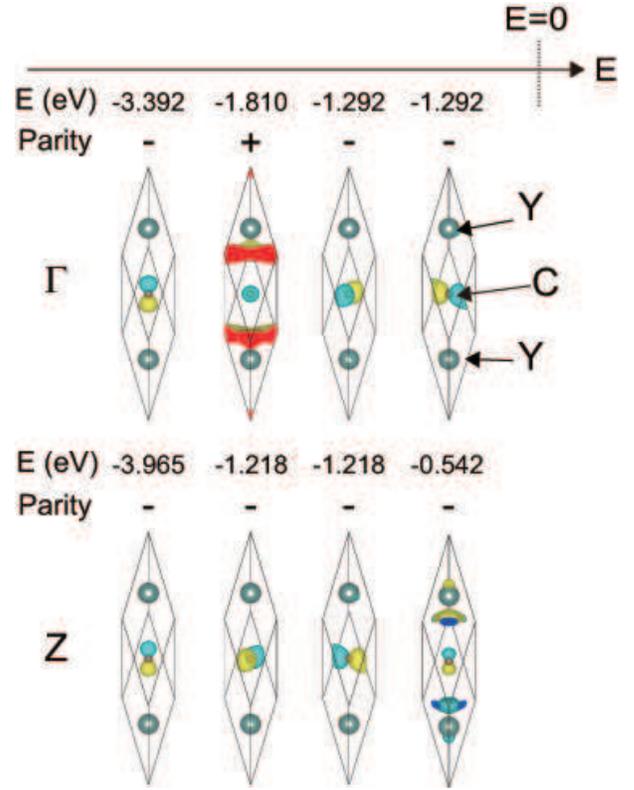}
\caption{\label{fig:parity} 
Parity eigenvalues in Y$_2$C.
Eigen energy, parity eigenvalues, and spatial distributions of eigenstates of occupied bands at the $\Gamma$ and $Z$ points of Y$_2$C in the GGA.  
The center of the unit cell is located at the C atom.
The energy is measured from the Fermi level.
}
\end{figure}

The parity eigenvalues of the four highest occupied eigenstates of Y$_2$C are shown in Fig.~\ref{fig:parity}.
Here the spin-orbit coupling is neglected, and each state is spin degenerate.
Here, we put the inversion center at the C atom; thereby, the unit cell is chosen to have the C atom at its center. 
The Zak phase at surface TRIM can be calculated from the products of the parity eigenvalues over the 
occupied states at the two corresponding bulk TRIM \cite{Hughes}.
In particular, the Zak phase $\theta(\bar{\Gamma})$ at $\bar{\Gamma}$ is related with the parity eigenvalues $\xi({\bf k})$
at the $\Gamma$ and the $Z$ points in the following way.
\begin{equation}
e^{i\theta(\bar{\Gamma})}=\xi(\Gamma)\xi(Z)=-1,
\label{eq:parity}
\end{equation}
meaning that $\theta(\bar{\Gamma})=\pi$ (mod $2\pi$), in agreement with Fig.~\ref{fig:Y2Csf}(c).

This nontrivial Zak phase, i.e.~the parity inversion between $\Gamma$ and $Z$,
means band inversion between $\Gamma$ and $Z$. 
Namely, the states along the $Z$-$\Gamma$-$Z$ line are topologically distinct
from those in the limit where the Y$_2$C layers are displaced far away from each other. 
In the present case,  the 
topologically nontrivial band structure along the $Z$-$\Gamma$-$Z$ line comes 
from the occupied interstitial states, as 
is analogous to the topologically nontrivial phase in the one-dimensional 
SSH model. 
The Wannier functions for the occupied interstitial 
states are located at the edge of the unit cell. Their locations are 
displaced from the inversion center by a half of the lattice constant, which corresponds to the $\pi$ Zak phase, 
because the Zak phase gives the center position of the Wannier function. 
These occupied interstitial states also account for the parity inversion between 
the $\Gamma$ and the $Z$ points.

As a result of this $\pi$ Zak phase, there should appear surface states at $\bar{\Gamma}$, similar to the midgap 
states in the SSH model, and it agrees with our surface-state calculation in Fig.~\ref{fig:Y2Csf}(a).
Our choice of the unit cell corresponds to the surface termination at the 
cleavage between the Y$_2$C layers. Therefore, from these reasons the 
existence of the topological surface states at the $\bar{\Gamma}$ point is
the hallmark of the topologically nontrivial band structure in the topological electride
Y$_2$C, and this feature is absent in the calculation in the previous work \cite{Huang18}.
In fact, the surface states in our result (Fig.~\ref{fig:Y2Csf}(a)) are quite different from those in  
the previous paper on Y$_2$C \cite{Huang18}. The surface states
in \cite{Huang18} appeas only close to the $\bar{M}$ points (called $\bar{F}$ in \cite{Huang18}), whereas the surface states  in 
our result (Fig.~\ref{fig:Y2Csf}(a)) cover almost the entire surface Brillouin zone. 
Between these two results, the regions of the  surface Brillouin zone with and without 
the surface states are interchanged between our paper and \cite{Huang18}, and we do not know the  reason 
for this disagreement.  

\begin{figure}[ptb]
\includegraphics[width=7cm]{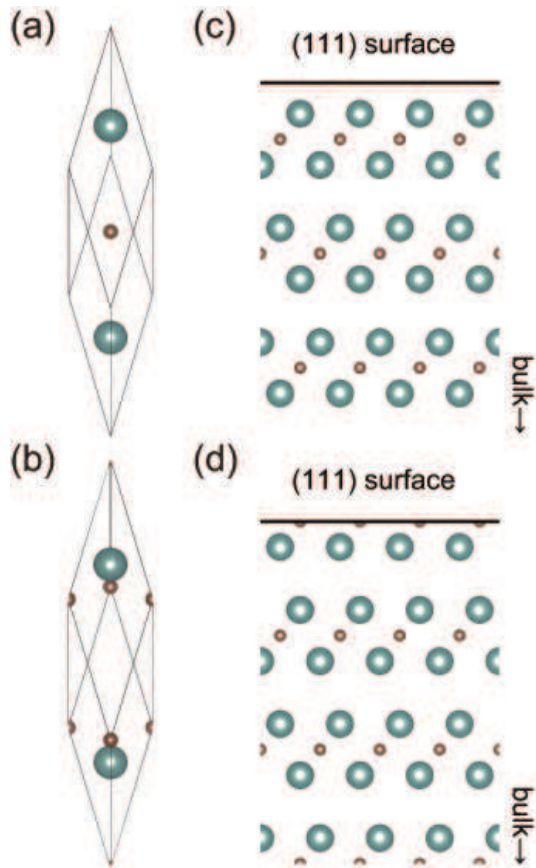}
\caption{\label{fig:uc} Two choices of the unit cell and corresponding 
surface terminations of Y$_2$C.
(a)(b) Two choices of the unit cell of Y$_2$C. The inversion centers 
are set at the centers of the unit cell in both cases. (a) is the one used in our calculation 
of the parity eigenvalues in the present paper, and the inversion 
center is at the carbon atom.
(b) shows another choice of the unit cell 
which is shifted by a half of 
the lattice constant from (a) along the [111] direction. 
(c) and (d) show the surface terminations of the (111) surface, 
which corresponds to the unit cell choices (a) and (b), respectively.
We note that in (d) the carbon atoms are cut in half, which is unphysical. 
}
\end{figure}

Next, we discuss the choices of the inversion center. In three-dimensional 
inversion-symmetric systems, there are eight choices for the inversion center,
and the parity eigenvalues depend on the choices. In the present paper, we take the inversion center to be
at the C atom (Fig.~\ref{fig:uc}(a)). Suppose we shift the inversion center 
along the [111] direction by a half of the lattice constant as shown in Fig.~\ref{fig:uc}(b), so that 
the inversion center is located in 
between the layers. Then the parity eigenvalues at $\Gamma$ are unchanged while those 
at $Z$ change their signs. Because the origin should be set at the inversion center, this change of the 
inversion center leads us to the change of the origin, leading to the 
gauge transformation for the periodic part of the Bloch wavefunction:
\begin{equation}
u_{\bf k}({\bf r})\ \rightarrow \ u'_{\bf k}({\bf r})=u_{\bf k}({\bf r})e^{ik_z a/2},
\end{equation}
where $a$ is the lattice constant along the [111] direction, and the $z$ axis 
is chosen along the [111] direction. This gauge transformation shifts the 
values of the Zak phase by $\pi$ per one electron (excluding the spin degeneracy).
When the total number of electrons divided by two (i.e. spin degeneracy) is odd,
this change will interchange the regions of the 0 Zak phase and that of the $\pi$ Zak phase.
In the present case, a half of the 
electron number is equal to 42, i.e. an even number, and the change of the inversion center
does not alter the Zak phase.

Physically, the change of the unit cell corresponds to the change of surface termination~\cite{Vanderbilt},
and the value of the Zak phase, being either $\pi$ or $0$ in the present case, corresponds to 
presence or absence of the midgap surface states. 
The two choices of the unit cell, Figs.~\ref{fig:uc} (a) and (b), correspond to
the surface terminations Figs.~\ref{fig:uc} (c) and (d), respectively. However, in Y$_2$C, the surface
termination shown in Fig.~\ref{fig:uc} (d) is not physical since it cuts carbon atoms into half,
and we do not study surface states with this surface termination.

\section{MOLECULES ON THE Sc$_2$C SURFACE}

\begin{figure*}[ptb]
\includegraphics[width=15cm]{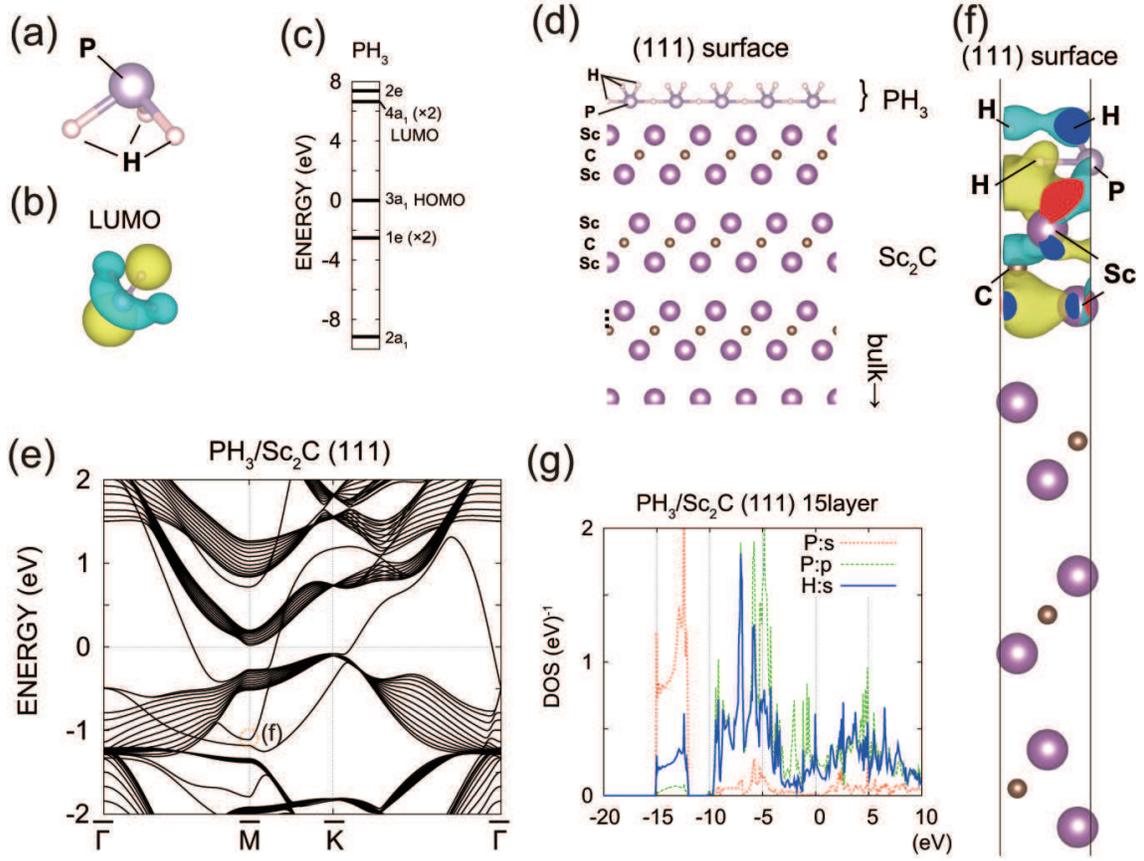}
\caption{\label{fig:mol} 
PH$_3$ on Sc$_2$C.
(a) Molecular structure of PH$_3$.
(b) One of the lowest unoccupied molecular orbital (LUMO) of PH$_3$.
(c) Eigen energy of PH$_3$.
(d) Crystal structure of PH$_3$ on the Sc$_2$C (111) surface.
(e) Electronic band structure of PH$_3$ on Sc$_2$C.
(f) Wavefunction of the eigenstate at the M point originating from PH$_3$.
The corresponding point is indicated by the dotted circle in (e).
(g) Partial density of states (pDOS) of PH$_3$ on the fifteen Sc$_2$C layers.
The energy is measured from the Fermi level.
}
\end{figure*}

We optimize the structure of various molecules on the (111) slab of the five Sc$_2$C layers.
Only in limited kinds of molecules, one can dope topological surface charges. 
Doping is impossible for molecules whose work functions are not small.
For example, in the case of the H$_2$O molecule on the Sc$_2$C, we find that the H$_2$O molecule does not absorb the topological surface charge and the molecule moves away from the surface.

Next, we shows the result of PH$_3$ on the Sc$_2$C slab.
The PH$_3$ molecule has the NH$_3$-type structure (Fig.~\ref{fig:mol}(a)).
PH$_3$ is a gas molecule at room temperature,
 and when the temperature is lowered, it passes through the liquid phase and becomes a solid phase under 140 K.
The lowest unoccupied molecular orbital (LUMO) of PH$_3$ originates from the H $s$ and the P $s$,$p$ orbitals (Figs.~\ref{fig:mol}(b,c)). 
The electronegativities of P and H are close, and the polarization of the molecule is not so large.
Because the work function of the LUMO is relatively deep, PH$_3$ is suitable for achieving $n$-type doping among various hydrogen compounds.
Indeed, we check the energy of the PH$_3$ molecule, the Sc$_2$C slab, and PH$_3$ on the both surface of the Sc$_2$C slab.
The energy of PH$_3$ on the Sc$_2$C slab is $0.3945 \times 2$ eV smaller than the total energy of two PH$_3$ molecules and the Sc$_2$C slab.
Figures~\ref{fig:mol}(d,e) show the structure of PH$_3$ on a 15-layer slab of Sc$_2$C and its electronic band structure, respectively.
The calculated charge distribution is shown in Fig.~\ref{fig:mol}(f).
We see that carriers are doped into the antibonding molecular orbitals of PH$_3$ originating from the H $s$ and P $s$,$p$ orbitals.
The anti-bonding orbitals are $p$-orbital-like molecular orbitals facing outward, 
and have a larger intermolecular hopping compared to the bonding orbitals.
The carrier doping to the anti-bonding orbitals somewhat destabilizes the molecules, but it does not break the structure.
Figure~\ref{fig:mol}(g) shows the partial density of states (pDOS) of PH$_3$ on the Sc$_2$C slab.
The distribution of the pDOS of the anti-bonding orbitals is broader than that of the bonding orbital,
since the anti-bonding orbital spreads outside the molecule.
In recent years, conventional superconductivity at 203 K was discovered in hydrogen sulfide at 155 GPa, recording the highest superconducting temperature to date~\cite{Li14,Drozdov15}.
In hydrogen sulfide, hydrogen carriers are generated at the Fermi level by metallization due to structural change to $Im\bar{3}m$ under high pressure.
On the other hand, in our case, by using the topological polarization and the shallowness of the work function of Sc$_2$C, we propose carrier doping directly to hydrogen without such a high pressure.
Unfortunately, the structure becomes metastable with the $1\times 2$ unit cell and the structure where a H$_2$ molecule is dissociated from the surface is most stable.
The design of the most stable structure of molecules with conductive hydrogen bands is a future problem.

\section{SYNTHESIS OF ELECTRIDE IN EXPERIMENT} 
In Table~\ref{table:ref}, 
we summarize a list of previous works on experimental 
synthesis of topological electrides studied in our paper.
We note that the previous experimental works on the synthesis of
Sc$_2$C reported cubic structure \cite{Rassaerts67,Sidorko95,Arellano11}, while 
theoretical works including our paper and Ref.~\cite{Zhang17} report
the trigonal structure  ($R\bar{3}m$) as the stable structure.
In experimental synthesis of Sc$_2$C,
one should be careful because
anionic hydrogen H$^-$ can easily replace
anionic electrons $e^-$. 
It might be the reason for the difference between the theoretical trigonal structure
and experimentally obtained cubic structure.

\begin{table}[tb] 
\caption{List of experimental synthesis of electride} 
\
\begin{tabular}{c|ccccc}
\hline \hline \\ [-8pt]  
Material               & Y$_2$C & Sr$_2$Bi & HfBr & LaBr   \\ [+1pt]
\hline \\ [-8pt] 
Ref.                 & \cite{Zhang14}  & \cite{Eisenmann14} & \cite{Marek79} & \cite{Mattausch80}    \\ 
\hline \hline 

\end{tabular}
\label{table:ref} 
\end{table}

\end{document}